\documentclass[aps,pra,preprint,showpacs,groupedaddress,amsmath,amssymb,amsfonts]{revtex4-1}
\usepackage{graphicx}
\usepackage{bm}

\begin{document}


\title{Quantum model for price forecasting in Financial Markets \\ ---------------{\footnotesize (bilingual english-spanish edition)}---------------\\
Modelo cu\'{a}ntico para la predicci\'{o}n de precios \\ en mercados financieros}


\author{J. L. Subias}
\email[e-mail for correspondence to the author:\\]{jlsubias@unizar.es}
\affiliation{Departamento de Ingenieria de Diseno y Fabricacion, Universidad de Zaragoza,\\C/Maria de Luna 3, 50018-Zaragoza, Spain}


\date{October 22, 2018}

\begin{abstract}
The present paper describes a practical example in which the probability distribution of the prices of a stock market blue chip is calculated as the wave function of a quantum particle confined in a potential well. This model may naturally explain the operation of several empirical rules used by technical analysts. Models based on the movement of a Brownian particle do not account for fundamental aspects of financial markets. This is due to the fact that the Brownian particle is a classical particle, while stock market prices behave more like quantum particles. When a classical particle meets an obstacle or a potential barrier, it may either bounce or overcome the obstacle,yet not both at a time. Only a quantum particle can simultaneously reflect and transmit itself on a potential barrier. This is precisely what prices in a stock market imitate when they find a resistance level: they partially bounce against and partially overcome it. This can only be explained by admitting that prices behave as quantum rather than as classic particles. The proposed quantum model finds natural justification not only for the aforementioned facts but also for other empirically well-known facts such as sudden changes in volatility, non-Gaussian distribution in prices, among others.
\newpage
{\large Resumen} \\
En el presente art\'{i}culo se describe un ejemplo pr\'{a}ctico en el que calculamos la distribuci\'{o}n de probabilidad de las cotizaciones de un blue-chip de un mercado burs\'{a}til como funci\'{o}n de onda de una part\'{i}cula cu\'{a}ntica confinada en un pozo de potencial. Este modelo podr\'{i}a explicar de forma natural el porqu\'{e} de varias reglas emp\'{i}ricas usadas por los analistas t\'{e}cnicos. Los modelos basados en el movimiento de una part\'{i}cula browniana dejan inexplicados aspectos fundamentales de los mercados financieros. Ello se debe a que la part\'{i}cula browniana es una part\'{i}cula cl\'{a}sica, mientras que las cotizaciones de los mercados se comportan m\'{a}s bien como part\'{i}culas cu\'{a}nticas. Cuando una part\'{i}cula cl\'{a}sica se encuentra con un obst\'{a}culo o barrera de potencial, puede rebotar o superar el obst\'{a}culo, pero no puede hacer ambas cosas a la vez. Solo si la part\'{i}cula es cu\'{a}ntica puede reflejarse y transmitirse simult\'{a}neamente ante una barrera de potencial. Eso es precisamente lo que remedan los precios en un mercado cuando encuentran un nivel de resistencia: en parte rebotan, en parte superan el nivel de resistencia. Esto solo es explicable admitiendo que los precios se comportan como part\'{i}culas cu\'{a}nticas m\'{a}s bien que cl\'{a}sicas. En el modelo cu\'{a}ntico que proponemos encuentran justificaci\'{o}n natural no solo los hechos anteriores, sino otros emp\'{i}ricamente bien conocidos, como los cambios bruscos de volatilidad, la distribuci\'{o}n no gaussiana de los precios y otros.
\end{abstract}


\maketitle

\section{Introduction}
Models extracted from statistical physics to describe financial market dynamics (Brownian models, pseudo Ising, etc.\cite{Dani03,Kaizo02,chow99}) have been followed by important works which make use of quantum mechanics to explain certain aspects neglected in previous works\cite{Raco01}. Quantum models have been proposed to consider the normative price-fluctuation limit imposed in several markets, which is equivalent to considering as a quantum system a particle in an infinite potential well or a harmonic oscillator with two extreme potential values\cite{CZhang01,Meng01}. Difficulty in these models lies in the estimation of the Hamiltonian of the stock market, which depends on such diverse factors as the economic situation, economic policies, market information, investor psychology, etc. The present model is aimed at overcoming this difficulty.
\section{Model description}
\begin{figure}[tb]
\centering
 \includegraphics*[scale=.7]{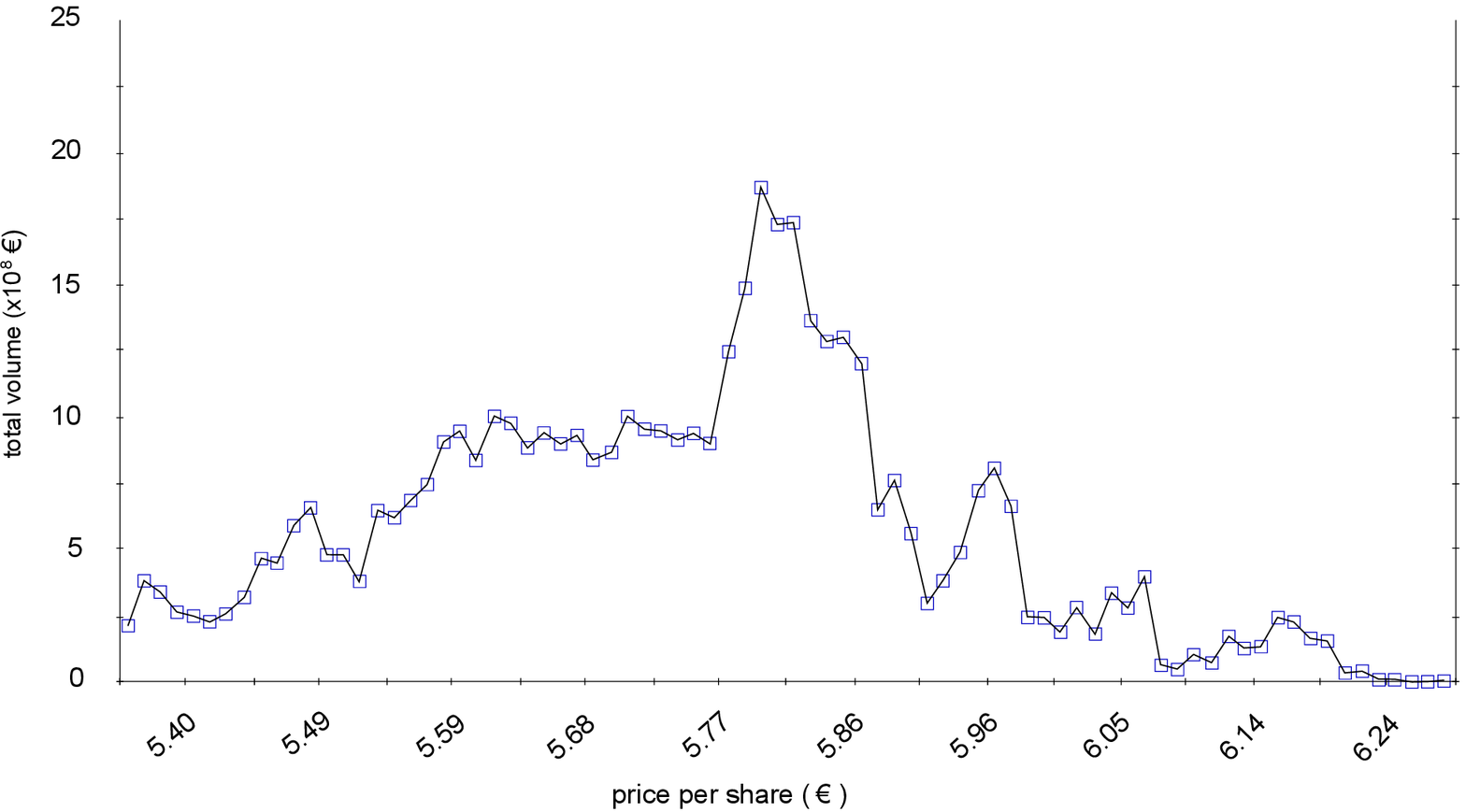}
  \caption{\label{campoPot}Potential field to which the particle is subjected, as calculated previously by statistical physics methods}
 \end{figure}

Let us assume that the stock market price of a blue chip behaves as a quantum particle with no spin confined in an infinite potential well. The bottom of the well forms potential $V(\mathfrak{p})$, which depends on recent past(see Fig.~\ref{campoPot}). The estimation of potential $V(\mathfrak{p})$ is a statistical physics problem which shall be tackled in a forthcoming paper.


The wave function which describes the state of the particle shall be conceived as a function $\Psi(\mathfrak{p},t)$ which depends on prices $\mathfrak{p}$ of the blue-chip shares and time $t$. In the space of states\footnote{\label{spc}Hilbert space of wave functions} and according to compact Dirac notation, we shall write:
\begin{equation}\label{ketPsi}
    |{\kern 1pt}\Psi\rangle=\sum_{n} c_{n}|{\kern 1pt}\phi_{n}\rangle,
\end{equation}
which indicates that the most general function $|{\kern 1pt}\Psi\rangle$ is a linear superposition of well-defined stationary wave functions, with complex coefficients $c_{n}=\langle\phi_{n}|{\kern 1pt}\Psi\rangle$.
Ket $|{\kern 1pt}\Psi\rangle$ would be biunivocally determined by the infinite set of $c_{n}$, $n=1,\ldots\infty$, this being its "price representation", which is formally analogous to representation $\{|\textbf{r}\rangle\}$ (positions) in quantum mechanics. The probability distribution of the blue-chip prices would be the squared module of the wave function -i.e., $|\Psi(\mathfrak{p},t|^{2}$.
\subsection{The Hamiltonian of a blue chip in a financial market}
In quantum mechanics there are no variables but observables, which are represented by Hermitian operators in a Hilbert space (space of states). In a financial market, each Hermitian operator would be represented by a set of economic or financial variables. Therefore, the next step is setting an equivalence between quantum-mechanical and financial concepts.
For the considered potential well, the Hamiltonian is:
\begin{equation}\label{hamiltonMerca}
    H=\frac{\mathbf{P}^{2}}{2m}+V,
\end{equation}
As mentioned above, Potential $V$ must be estimated by statistical physics procedures and, for the moment, let us assume it is known to us. Particle mass $m$ shall be related to the average daily trading volume expressed in thousand Euro. If the position of the particle is related to blue-chip stock price $\mathfrak{p}$ in time $t$, consequently, linear momentum $\mathbf{P}$ shall be related to the variation rate of this price throughout time.
\subsection{The Schr\"{o}dinger equation}
In this situation, the Schr\"{o}dinger equation which rules the evolution of the particle would be, in its compact form: $H|\phi\rangle=E|\phi\rangle$, and explicitly:
\begin{equation}\label{SchroIndepExpli}
    \frac{-\hbar^{2}}{2m}\frac{\partial^{2}\phi_{n}}{\partial x^{2}}+V(\mathfrak{p})\phi_{n}=E_{n}\phi_{n},
\end{equation}
which is the time-independent Schr\"{o}dinger equation, where $\phi_{n}$ is the well-defined energy stationary wave function $E_{n}$.
The time-dependent Schr\"{o}dinger equation in its compact expression would be:
\begin{equation}\label{SchroDep}
    H|\Psi(\mathfrak{p},t)\rangle=i\hbar^{2}\frac{\partial}{\partial t}|\Psi\rangle,
\end{equation}
And, explicitly:
\begin{equation}\label{SchroDepExpli}
    \frac{-\hbar^{2}}{2m}\frac{\partial^{2}}{\partial \mathfrak{p}^{2}}\Psi+V(\mathfrak{p})\Psi=i\hbar\frac{\partial}{\partial t}\Psi,
\end{equation}
whose general solution is
\begin{equation}\label{SchroDepExpliSolu}
    \Psi(\mathfrak{p},t)=\sum_{n}c_{n}\phi_{n}e^{-i\frac{E_{n}}{\hbar}t},
\end{equation}
That is, the general solution $\Psi(\mathfrak{p},t)$ is a linear combination of an infinite number of stationary wave functions $\phi_{n}$, $n=1,\ldots\infty$.
\subsection{Thermodynamic equilibrium and average energy}
 Assuming that the system is in equilibrium with a thermal bath at temperature $T$,the state of the system could be described by a statistical ensemble of stationary states $|\phi_{n}\rangle$ with weighing coefficients proportional to $\exp(-E_{n}/(KT))$, where $E_{n}$ is the energy in state $|\phi_{n}\rangle$, $K$ is the Boltzmann constant, and  $T$ is the equilibrium temperature. This mixture state can be characterized by the density matrix, which can be described in this case as:
\begin{equation}\label{matrizDensi}
    \rho=\frac{1}{Z}e^{\frac{-H}{KT}},
\end{equation}
where $H$ is the Hamiltonian operator, and $Z$ is the canonical partition function, shaped as:
\begin{equation}\label{funcParti}
    Z=Tr\{e^{\frac{-H}{KT}}\},
\end{equation}
Or more explicitly:
\begin{equation}\label{funcPartiExpli}
    Z=\sum^{\infty}_{n=1}\langle\phi_{n}|e^{\frac{-H}{KT}}|\phi_{n}\rangle=\sum^{\infty}_{n=1}e^{\frac{-E_{n}}{KT}},
\end{equation}
where $E_{n}$ are the eigenvalues of Hamiltonian $H$.
 In the corresponding eigenvector base $\{|\phi_{n}\rangle\}$ the matrix elements of the  $\rho$ density operator can be expressed as:
\begin{equation}\label{Ronn}
    \rho_{nn}=\frac{1}{Z}\langle\phi_{n}|e^{\frac{-H}{KT}}|\phi_{n}\rangle=\frac{1}{Z}e^{\frac{-E_{n}}{KT}},
\end{equation}
\begin{equation}\label{Ronm}
    \rho_{nm}=\langle\phi_{n}|e^{\frac{-H}{KT}}|\phi_{m}\rangle=\frac{1}{Z}e^{\frac{-E_{n}}{KT}}\langle\phi_{n}|\phi_{m}\rangle=0 ,
\end{equation}
 Relationships ~(\ref{Ronn}), ~(\ref{Ronm}) mean that -in thermodynamic equilibrium- eigenstate populations exponentially decrease with energy as well as that $\rho_{nm}$ coherence between eigenstates is zero.
\subsubsection{Two-level quantum-system approximation}
\begin{figure}[tb]
\centering
 \includegraphics*[scale=.7]{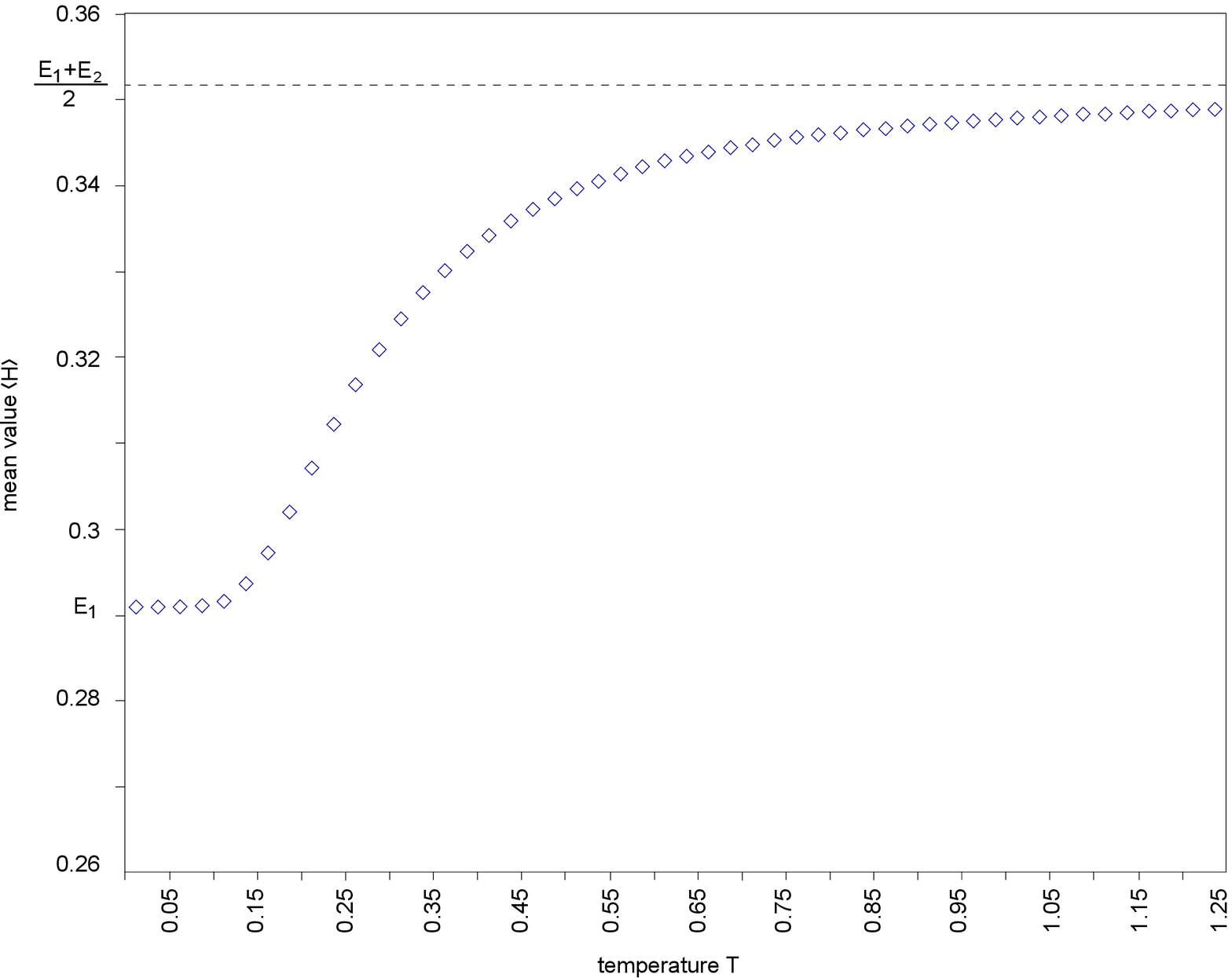}
  \caption{\label{HversusT}Average value of Hamiltonian $\langle H \rangle$ vs. absolute temperature $T$ for a two-level $E_{1}$, $E_{2}$ quantum system}
 \end{figure}

Using the $\rho$ density matrix, average value $\langle H \rangle$ can be calculated as:
\begin{equation}\label{Hmed}
    \langle H \rangle=Tr(H \rho)=\frac{1}{Z}Tr(He^{\frac{-H}{KT}}),
\end{equation}
Expanding this expression in eigenvector base $\{\phi_{n}\}$ it can be seen that:
\begin{equation}\label{HmedExpan}
    \langle H \rangle=\frac{1}{Z}\sum_{n=1}^{\infty}E_{n}e^{\frac{-E_{n}}{KT}},
\end{equation}
At this point, we shall introduce a relevant simplification: empirical data show that -for the quantum particle- only the ground state and the first excited state are reachable, since the second and following excited states demand increasing energy levels which are unreachable for the particle. That is, the problem is reduced to a two-state (fundamental and first excited) system. This reduction is based on the fact that a stock price can suddenly pass from low to high volatility to then go back to its previous low volatility state. This fact has not been explained so far, being tackled only in \textit{ad hoc} models. For this two-level system, the general equation ~(\ref{HmedExpan}) renders the following result:
\begin{equation}\label{Hmed2niv}
    \langle H \rangle=\frac{E_{1}e^{\frac{-E_{1}}{KT}}+E_{2}e^{\frac{-E_{2}}{KT}}}{e^{\frac{-E_{1}}{KT}}+e^{\frac{-E_{2}}{KT}}},
\end{equation}
$E_{1}$, $E_{2}$ being well-defined energy levels, $K$ being the Boltzmann constant, $T$ being absolute temperature. Fig.~\ref{HversusT} is a graph as a function of $T$ which shows that $\langle H \rangle$ approaches to $E_{1}$ when temperature $T$ approaches to zero, and asymptotically to $(E_{1}+E_{2})/2$ when $T$ reaches high values.


The meaning of Fig.~\ref{HversusT} is easy to understand: for $T=0$,the system is in its ground state $|\Psi_{1}\rangle$ of energy $E_{1}$.At high temperatures, the statistical mixture is formed by both levels $E_{1}$ and $E_{2}$ equally, and average $\langle H \rangle$ approximates one half of the sum of $E_{1}$ and $E_{2}$.
Applying formulae ~(\ref{Ronn}), ~(\ref{Ronm}) the elements of the density matrix for the system of two energy levels are obtained:
\begin{eqnarray}\label{Ro2niv}
   \rho_{1,1}=\ldots=\frac{e^{\frac{-E_{1}}{KT}}}{e^{\frac{-E_{1}}{KT}}+e^{\frac{-E_{2}}{KT}}},
   \hspace{2cm} \rho_{1,2}=\rho_{2,1}=0 \\
   \nonumber \rho_{2,2}=\ldots=\frac{e^{\frac{-E_{2}}{KT}}}{e^{\frac{-E_{1}}{KT}}+e^{\frac{-E_{2}}{KT}}}
\end{eqnarray}
\subsection{Market temperature}
Assuming that prices $\mathfrak{p}$, , rates of change (returns) $\mathfrak{r}$, and trading volumes $\mathfrak{v}$ play the roles of particle position, speed and mass, respectively, the uncertainty principle can be expressed as:
\begin{equation}\label{incerti}
    \triangle \mathfrak{p}\triangle (\mathfrak{v}\mathfrak{r})\geq\hbar/2,
\end{equation}
formally being: $\mathfrak{v}\mathfrak{r}= momentum$, and $\mathfrak{r}=\text{d}\mathfrak{p}/\text{d}t$.
Kinetic energy $E_{k}$ averaged over the time would be:
\begin{equation}\label{Ecine}
    \langle E_{k} \rangle=\langle \frac{(\mathfrak{vr})^{2}}{2\mathfrak{v}} \rangle=\frac{1}{2}KT,
\end{equation}
where $K$ is the Boltzmann.
Consequently, a coherent definition of temperature would be:
\begin{equation}\label{T}
    T=\frac{\langle\mathfrak{v}\mathfrak{r}^{2}\rangle}{K},
\end{equation}
For further discussion on the concept of market temperature, see \cite{Subi13}.
\subsection{Probability distribution of the observable \textit{price}}
Now, let us find probability $\rho(\mathfrak{p})\text{d}\mathfrak{p}$ of finding a particle in a position $\mathfrak{p}$ located between $\mathfrak{p}$ and $\mathfrak{p}+\text{d}\mathfrak{p}$. When the particle is in stationary state $|\phi_{n}\rangle$,the corresponding probability density $\rho_{n}(\mathfrak{p})$ will be:
\begin{equation}\label{formRon}
    \rho_{n}(\mathfrak{p})=|\phi_{n}(\mathfrak{p})|^{2}=\langle\mathfrak{p}|\phi_{n}\rangle\langle\phi_{n}|\mathfrak{p}\rangle,
\end{equation}
In thermodynamic equilibrium, particle state is described by a statistical ensemble of states $|\phi_{n}\rangle$ with weighing coefficients $(1/Z)\exp(-En/(KT))$. Therefore, probability density in this case will be:
\begin{equation}\label{formRo}
    \rho(\mathfrak{p})=\frac{1}{Z}\sum_{n}\rho_{n}(\mathfrak{p})e^{\frac{-E_{n}}{KT}},
\end{equation}
 That is, probability density $\rho(\mathfrak{p})$ can be defined as the weighed sum of densities $\rho_{n}(\mathfrak{p})$ corresponding to the diverse states $|\phi_{n}\rangle$.
Let us see how the probability density defined in ~(\ref{formRo}) is related to density matrix $\rho$. Combining ~(\ref{formRo}) and ~(\ref{formRon}) renders:
\begin{equation}\label{rop}
    \rho(\mathfrak{p})=\frac{1}{Z}\sum_{n}\rho_{n}(\mathfrak{p})e^{\frac{-E_{n}}{KT}}\langle\mathfrak{p}|\phi_{n}\rangle\langle\phi_{n}|\mathfrak{p}\rangle,
\end{equation}
Bearing in mind the closure relation for states $|\phi_{n}\rangle$, operator $\exp(-H/(KT))$ can be rewritten in the following way:
\begin{equation}\label{eH}
    e^{\frac{-H}{KT}}=e^{\frac{-H}{KT}}\sum_{n}|\phi_{n}\rangle\langle\phi_{n}|=\sum_{n}e^{\frac{-E_{n}}{KT}}|\phi_{n}\rangle\langle\phi_{n}|,
\end{equation}
Combining ~(\ref{rop}) and ~(\ref{eH}):
\begin{equation}\label{G}
    \rho(\mathfrak{p})=\frac{1}{Z}\langle \mathfrak{p}|e^{\frac{-H}{KT}}|\mathfrak{p}\rangle=\langle \mathfrak{p}|\rho|\mathfrak{p}\rangle,
\end{equation}
Thus, it can be interpreted that $\rho(\mathfrak{p})$ is rho's diagonal element corresponding to ket $|{\kern 1pt}\mathfrak{p}\rangle$.
\section{Step-by-step description of a practical case}
Next, the probability distribution of the prices of a blue chip in Spanish IBEX 35 stock market shall be calculated -as well as the steps to follow shall be described.
\subsection{Setting up boundary conditions}
First of all, the time period through which we want the obtained result to be valid (probability distribution) shall be set up. The walls of the well determine an infinite potential -i.e., the quantum particle cannot flee the well in which it is confined. This assumption is based on the empirical evidence that the price of any market cap -be it blue chip or not- cannot statistically exceed a maximum exchange rate which depends on the considered time period. Thus, for instance, a market cap is rather unlikely -save for a stock market crash- to exceed $30$\% within a period of $20$ trading sessions. This limitation is equivalent to the walls of the well. Consequently, analysis of the blue chip's historical prices allows us to determine the maximum price fluctuation which may take place in the observed time period with a $95$\% confidence level, except for stock market innovations and crashes. In this case, a period of  $30$ days was set, which determined $\pm15$\% in maximum price fluctuation. This means that the walls of the well shall be placed $\pm15$\% from the last price which occupied the central position.


\begin{figure}[tb]
\centering
 \includegraphics*[scale=.7]{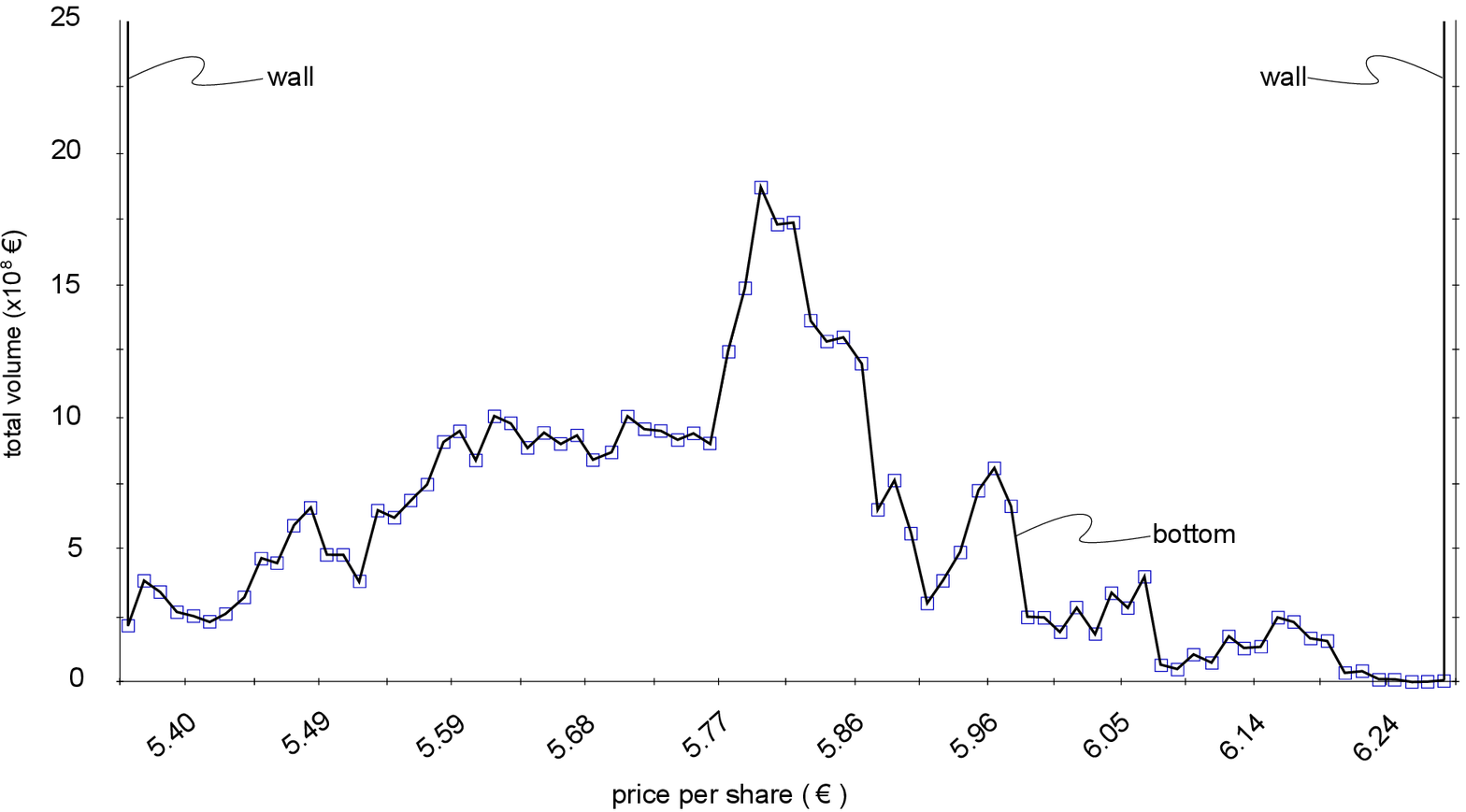}
  \caption{\label{pozoPot}The potential field, \textit{a priori} calculated, shapes the bottom of the potential well in which the quantum particle is confined}
 \end{figure}
\subsection{Shaping the bottom of the well}
The application of statistical physics methods allows us to estimate function $V(\mathfrak{p})$ ), which shall in general show peaks and valleys depending on recent market history. As mentioned previously, the description of this method shall be tackled in a forthcoming paper. Figure ~\ref{pozoPot} shows the potential well and function $V(\mathfrak{p})$ as estimated on October 2nd, 2017. This function $V(\mathfrak{p})$ gradually changes throughout time. However, it can be considered constant for a period of $20$ trading days, save for sudden bursts of innovation.
\subsection{Solving the Schr\"{o}dinger equation}
Function $V(\mathfrak{p})$ shall be discretized on $N=100$ points separated by constant interval  $\triangle=\mathfrak{p}_{i+1}-\mathfrak{p}_{i}$. This discretization process and the application of the finite difference method allow us to reduce the Schr\"{o}dinger equation to a homogeneous system of $N$ linear algebraic equations which, in short, is a problem of eigenvalues in matrix
\[
\left(
  \begin{array}{ccccc}
    v_{1} & -1 & 0 & \ldots & 0 \\
    -1 & v_{2} & -1 &  & 0 \\
     0& -1 & \ddots & \ddots & \vdots \\
     \vdots&  & \ddots & v_{n-1} & -1 \\
     0& 0 & \ldots & -1 & v_{n} \\
  \end{array}
\right)
\]
 where $v_{i}=2+\triangle^{2}2\mathfrak{v}V(\mathfrak{p}_{i})/\hbar^{2}$
 can be quickly and easily diagonalized, thus solving the problem. A standard implementation needs only setting up the appropriate ratio conversion factors of price returns to distance in angstrom {\AA} and of total trading volume to energy in electronvolt.
\subsection{Calculating the elements in the density matrix}
The previous section shall have provided us with energy levels $E_{1}, E_{2},\ldots, E_{n}$, as well as with stationary wave functions $\phi_{1}, \phi_{2},\ldots, \phi_{n}$ (see Fig.~\ref{FO1}). The temperature calculated from volatility with formula~(\ref{T}) -together with energy levels $E_{1}$, $E_{2}$- shall provide us with $\rho_{1,1}$, $\rho_{2,2}$ from formulae~(\ref{Ronn}), ~(\ref{Ronm}).
\subsection{Calculating the price probability distribution}
\begin{figure}[tb]
\centering
 \includegraphics*[scale=.7]{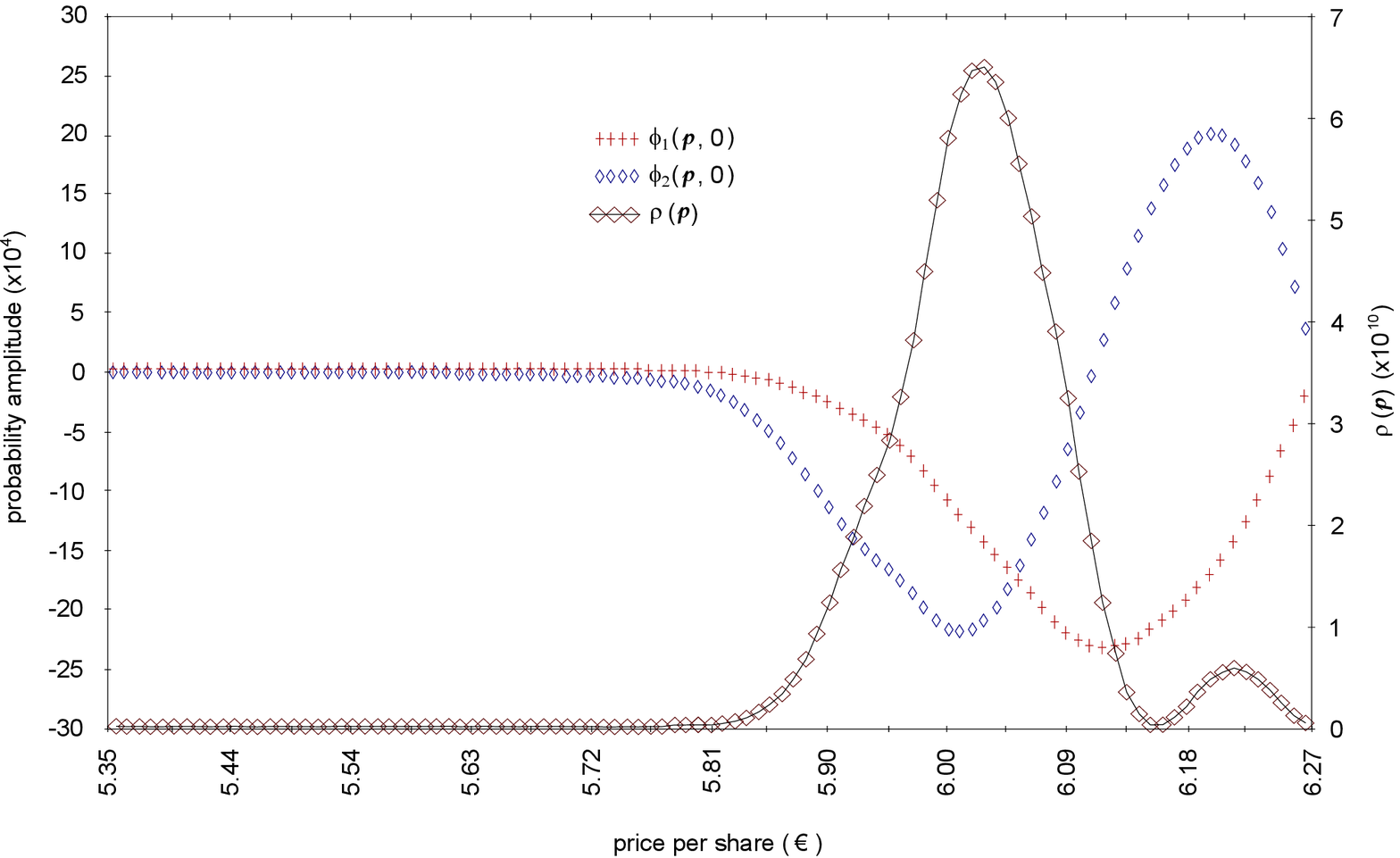}
  \caption{\label{FO1}Wave functions corresponding to both energy levels and probability distribution of the observable \textit{price}}
 \end{figure}

Final solution shall be function $\rho(\mathfrak{p})$ calculated from ~(\ref{G}) -see Fig.~\ref{FO1}.


\section{Conclusions}
The present paper presents a model to describe financial market dynamics based on the time evolution of a particle placed in a potential field -more precisely, a quantum particle confined in an infinite potential well. Its key features are described next:
\begin{itemize}
  \item 1)	Regarding price fluctuation, the statistical limit existing \textit{de facto} in all markets is considered, instead of the legal-normative level considered in other models, and
  \item 2)	An infinite potential well is considered. Its bottom is not flat but a function of recent market history, instead of the infinite square well, the periodical external field or the harmonic potential considered in other models.
\end{itemize}
According to our experience, this model proves coherent and capable of explaining certain aspects of financial markets which remain unexplained by models based on classic (Brownian) particles. This model allows obtaining probability distributions of prices for market blue chips, as well as being used with forecasting aims.

\newpage
\framebox{Spanish version}

\section{Introducci\'{o}n}
Con posterioridad a los modelos entresacados de la F\'{i}sica Estad\'{i}stica para describir la din\'{a}mica de los mercados financieros (modelos brownianos, pseudo-Ising, etc\cite{Dani03,Kaizo02,chow99}) ha habido importantes trabajos que involucran la Mec\'{a}nica Cu\'{a}ntica para explicar aspectos no descritos por los anteriores modelos\cite{Raco01}. Se han propuesto modelos cu\'{a}nticos que consideran el limite normativo de fluctuaci\'{o}n de precios impuesto en algunos mercados, lo cual equivale a considerar como sistema cu\'{a}ntico una part\'{i}cula dentro de un pozo de potencial infinito o un oscilador arm\'{o}nico con dos valores extremos de potencial\cite{CZhang01,Meng01}. La dificultad de estos modelos reside en la estimaci\'{o}n del hamiltoniano del mercado financiero, que depende de factores tan diversos como la coyuntura econ\'{o}mica, las pol\'{i}ticas econ\'{o}micas, la informaci\'{o}n del mercado, la psicolog\'{i}a de los inversores, etc. Esta es la dificultad que pretendemos superar con el presente modelo.
\section{Desdripci\'{o}n del modelo}
\begin{figure}[tb]
\centering
 \includegraphics*[scale=.7]{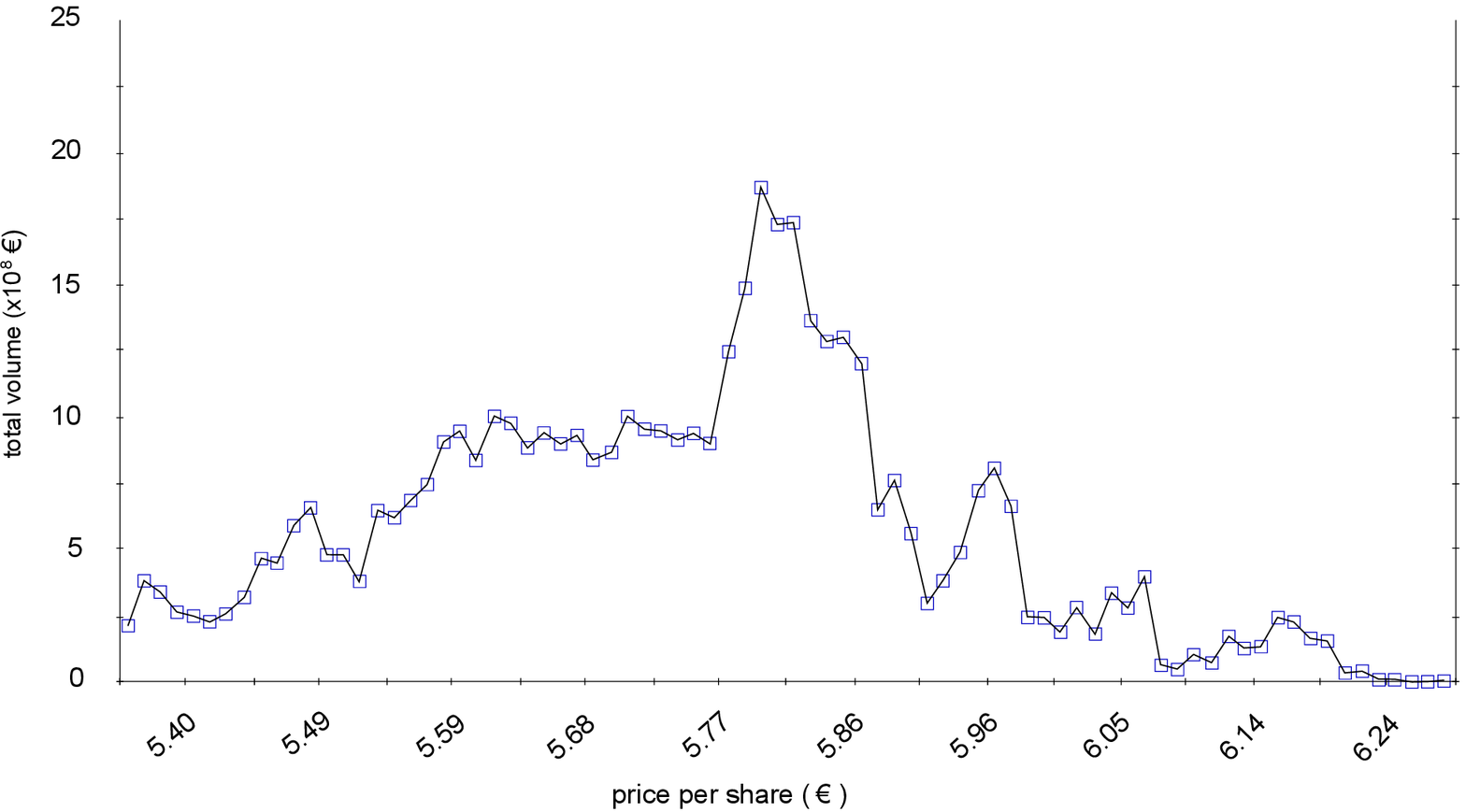}
  \caption{\label{campoPotS}Campo de potencial al que est\'{a} sometida la particula. Se calcula previamente por metodos de F\'{i}sica Estad\'{i}stica}
 \end{figure}

Asumiremos que la cotizaci\'{o}n de un stock suficientemente capitalizado (blue-chip) se comporta como una part\'{i}cula cu\'{a}ntica sin spin encerrada en un pozo de potencial infinito. El fondo del pozo conforma un potencial $V(\mathfrak{p})$ que depende del pasado reciente (ver Fig.~\ref{campoPotS}). La estimaci\'{o}n de este potencial $V(\mathfrak{p})$ constituye un problema de F\'{i}sica Estad\'{i}stica al cual dedicaremos un futuro art\'{i}culo.


La funci\'{o}n de onda que describe el estado de la part\'{i}cula la concebiremos como una funci\'{o}n $\Psi(\mathfrak{p},t)$ que depende de las cotizaciones $\mathfrak{p}$ de las acciones del blue-chip y del tiempo $t$. En el espacio de estados\footnote{\label{spcS}Espacio de Hilbert de las funciones de onda} y adoptando la notaci\'{o}n compacta de Dirac, escribiremos:
\begin{equation}
    |{\kern 1pt}\Psi\rangle=\sum_{n} c_{n}|{\kern 1pt}\phi_{n}\rangle,
\end{equation}
Que indica que la funci\'{o}n m\'{a}s general $|{\kern 1pt}\Psi\rangle$ es una superposici\'{o}n lineal de funciones de onda $|{\kern 1pt}\phi_{n}\rangle$ estacionarias de energ\'{i}a bien definida, con coeficientes complejos $c_{n}=\langle\phi_{n}|{\kern 1pt}\Psi\rangle$.
El ket $|{\kern 1pt}\Psi\rangle$ quedar\'{i}a biun\'{i}vocamente determinado por el conjunto infinito de los $c_{n}$, $n=1,\ldots\infty$, siendo esta su "representaci\'{o}n-cotizaciones" , formalmente an\'{a}loga a la representaci\'{o}n $\{|\textbf{r}\rangle\}$ (posiciones) de la Mec\'{a}nica Cu\'{a}ntica. La distribuci\'{o}n de probabilidad de las cotizaciones del blue-chip ser\'{i}a el m\'{o}dulo al cuadrado de la funci\'{o}n de onda, es decir, $|\Psi(\mathfrak{p},t|^{2}$.
\subsection{Hamiltoniano de un blue-chip en un mercado financiero}
En Mec\'{a}nica Cu\'{a}ntica no hay variables, sino observables, que vienen representados por operadores hermiticos en un espacio de Hilbert (espacio de estados). En un mercado financiero cada operador hermitico estar\'{i}a representado por un conjunto de variables econ\'{o}micas o financieras. Por tanto, el siguiente paso que daremos ser\'{a} establecer una equivalencia entre conceptos mecano-cu\'{a}nticos y financieros.
Para el pozo de potencial que estamos considerando, el hamiltoniano adopta la forma:
\begin{equation}
    H=\frac{\mathbf{P}^{2}}{2m}+V,
\end{equation}
El potencial $V$, como ya hemos dicho, hay que estimarlo por procedimientos de F\'{i}sica Estad\'{i}stica y, de momento, lo asumiremos conocido. La masa $m$ de la part\'{i}cula la asimilaremos al volumen medio de negociaci\'{o}n diaria expresado en miles de euros. Si la posici\'{o}n de la part\'{i}cula la asimilamos a la cotizaci\'{o}n $\mathfrak{p}$ del blue-chip en un instante $t$ del tiempo, consecuentemente, el momento lineal $\mathbf{P}$ se relacionar\'{a} con la tasa de variaci\'{o}n de esa cotizaci\'{o}n a lo largo del tiempo.
\subsection{Ecuaci\'{o}n de Schr\"{o}dinger}
En esta situaci\'{o}n la ecuaci\'{o}n de Schr\"{o}dinger que gobierna la evoluci\'{o}n de la part\'{i}cula ser\'{i}a, en forma compacta: $H|\phi\rangle=E|\phi\rangle$, y expl\'{i}citamente:
\begin{equation}
    \frac{-\hbar^{2}}{2m}\frac{\partial^{2}\phi_{n}}{\partial x^{2}}+V(\mathfrak{p})\phi_{n}=E_{n}\phi_{n},
\end{equation}
Que es la ecuaci\'{o}n de Schr\"{o}dinger independiente del tiempo, donde $\phi_{n}$ es la funci\'{o}n de onda estacionaria de energ\'{i}a $E_{n}$ bien definida.
La ecuaci\'{o}n de Schr\"{o}dinger dependiente del tiempo ser\'{i}a en notaci\'{o}n compacta:
\begin{equation}
    H|\Psi(\mathfrak{p},t)\rangle=i\hbar^{2}\frac{\partial}{\partial t}|\Psi\rangle,
\end{equation}
Y expl\'{i}citamente:
\begin{equation}
    \frac{-\hbar^{2}}{2m}\frac{\partial^{2}}{\partial \mathfrak{p}^{2}}\Psi+V(\mathfrak{p})\Psi=i\hbar\frac{\partial}{\partial t}\Psi,
\end{equation}
Cuya soluci\'{o}n general es
\begin{equation}
    \Psi(\mathfrak{p},t)=\sum_{n}c_{n}\phi_{n}e^{-i\frac{E_{n}}{\hbar}t},
\end{equation}
Es decir, la soluci\'{o}n general $\Psi(\mathfrak{p},t)$ es una combinaci\'{o}n lineal de un n\'{u}mero infinito de funciones de onda estacionarias $\phi_{n}$, $n=1,\ldots\infty$.
\subsection{Equilibrio termodin\'{a}mico y valor medio de la energ\'{i}a}
Asumiendo que el sistema se encuentre en equilibrio con un ba\~{n}o t\'{e}rmico a temperatura $T$, el estado del sistema podr\'{i}a ser descrito por una colectividad estad\'{i}stica de estados estacionarios $|\phi_{n}\rangle$ con coeficientes de ponderaci\'{o}n proporcionales a $\exp(-E_{n}/(KT))$, donde $E_{n}$ es la energ\'{i}a del estado $|\phi_{n}\rangle$, $K$ la constante de Boltzmann y $T$ la temperatura de equilibrio. Tal estado mezcla se puede caracterizar por la matriz densidad, que en este caso se puede escribir como
\begin{equation}
    \rho=\frac{1}{Z}e^{\frac{-H}{KT}},
\end{equation}
Donde $H$ es el operador hamiltoniano y $Z$ la funci\'{o}n de partici\'{o}n can\'{o}nica, que adopta la forma
\begin{equation}
    Z=Tr\{e^{\frac{-H}{KT}}\},
\end{equation}
O m\'{a}s expl\'{i}citamente
\begin{equation}
    Z=\sum^{\infty}_{n=1}\langle\phi_{n}|e^{\frac{-H}{KT}}|\phi_{n}\rangle=\sum^{\infty}_{n=1}e^{\frac{-E_{n}}{KT}},
\end{equation}
Donde los $E_{n}$ son los valores propios del hamiltoniano $H$.
En la correspondiente base $\{|\phi_{n}\rangle\}$ de vectores propios los elementos de matriz del operador densidad $\rho$ se pueden escribir como
\begin{equation}\label{RonnS}
    \rho_{nn}=\frac{1}{Z}\langle\phi_{n}|e^{\frac{-H}{KT}}|\phi_{n}\rangle=\frac{1}{Z}e^{\frac{-E_{n}}{KT}},
\end{equation}
\begin{equation}\label{RonmS}
    \rho_{nm}=\langle\phi_{n}|e^{\frac{-H}{KT}}|\phi_{m}\rangle=\frac{1}{Z}e^{\frac{-E_{n}}{KT}}\langle\phi_{n}|\phi_{m}\rangle=0 ,
\end{equation}
Las anteriores relaciones ~(\ref{RonnS}), ~(\ref{RonmS}) significan que, en equilibrio termodin\'{a}mico, las poblaciones de estados propios son exponencialmente decrecientes con la energ\'{i}a y que las coherencias $\rho_{nm}$ entre estados propios son cero.
\subsubsection{Aproximaci\'{o}n de sistema cu\'{a}ntico de dos niveles}
\begin{figure}[tb]
\centering
 \includegraphics*[scale=.7]{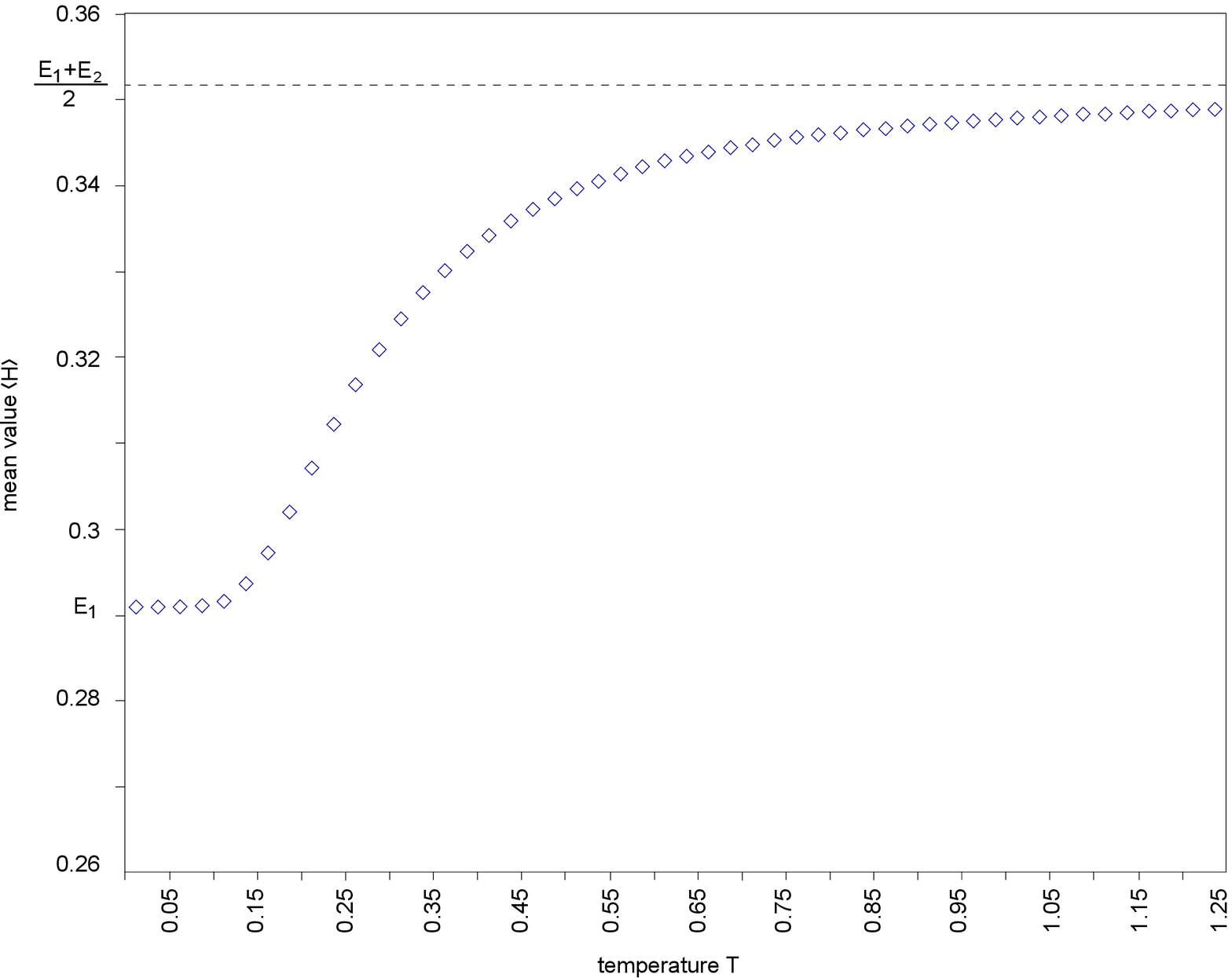}
  \caption{\label{HversusTs}Valor medio del hamiltoniano $\langle H \rangle$ versus temperatura absoluta $T$ para un sistema cuantico de dos niveles $E_{1}$, $E_{2}$ }
 \end{figure}

Usando la matriz de densidad $\rho$ se puede calcular el valor medio $\langle H \rangle$ :
\begin{equation}
    \langle H \rangle=Tr(H \rho)=\frac{1}{Z}Tr(He^{\frac{-H}{KT}}),
\end{equation}
Expandiendo esta expresi\'{o}n en la base de vectores propios $\{\phi_{n}\}$ tenemos:
\begin{equation}\label{HmedExpanS}
    \langle H \rangle=\frac{1}{Z}\sum_{n=1}^{\infty}E_{n}e^{\frac{-E_{n}}{KT}},
\end{equation}
En este punto vamos a introducir una importante simplificaci\'{o}n: en base a datos emp\'{i}ricos, asumiremos que, para la part\'{i}cula cu\'{a}ntica, solo son alcanzables el estado fundamental y el primer excitado, debido a que el segundo excitado y siguientes requieren niveles de energ\'{i}a crecientes e inalcanzables para la part\'{i}cula. Es decir, reducimos el problema a un sistema cu\'{a}ntico de dos estados (fundamental y primer excitado). Esta asunci\'{o}n est\'{a} basada en el hecho de que un mercado burs\'{a}til puede pasar s\'{u}bitamente de baja a alta volatilidad para, posteriormente, retornar a su estado anterior de baja volatilidad. Este hecho, hasta ahora, no ha sido explicado, sino solamente simulado por modelos "ad hoc". Para tal sistema de dos niveles la ecuaci\'{o}n general ~(\ref{HmedExpanS}) da como resultado:
\begin{equation}
    \langle H \rangle=\frac{E_{1}e^{\frac{-E_{1}}{KT}}+E_{2}e^{\frac{-E_{2}}{KT}}}{e^{\frac{-E_{1}}{KT}}+e^{\frac{-E_{2}}{KT}}},
\end{equation}
Siendo $E_{1}$, $E_{2}$ los niveles de energ\'{i}a bien definidos, $K$ la constante de Bolzmann, $T$ la temperatura absoluta. La grafica en funci\'{o}n de $T$ puede verse en Fig.~\ref{HversusTs}, en la cual se aprecia que cuando la temperatura $T$ tiende a cero, $\langle H \rangle$ tiende a $E_{1}$ y cuando $T$ crece hasta valores altos, $\langle H \rangle$ tiende asint\'{o}ticamente a $(E_{1}+E_{2})/2$


El significado de Fig.~\ref{HversusTs} es f\'{a}cil de entender: para $T=0$ el sistema se encuentra en su estado fundamental $|\Psi_{1}\rangle$ de energ\'{i}a $E_{1}$; a altas temperaturas la mezcla estad\'{i}stica est\'{a} formada por ambos niveles $E_{1}$, $E_{2}$ a partes iguales, y el promedio $\langle H \rangle$ se aproxima a la semisuma de $E_{1}$ y $E_{2}$.
Aplicando las f\'{o}rmulas ~(\ref{RonnS}), ~(\ref{RonmS}) obtenemos los elementos de la matriz densidad para el sistema de dos niveles de energ\'{i}a:
\begin{eqnarray}
   \rho_{1,1}=\ldots=\frac{e^{\frac{-E_{1}}{KT}}}{e^{\frac{-E_{1}}{KT}}+e^{\frac{-E_{2}}{KT}}},
   \hspace{2cm} \rho_{1,2}=\rho_{2,1}=0 \\
   \nonumber \rho_{2,2}=\ldots=\frac{e^{\frac{-E_{2}}{KT}}}{e^{\frac{-E_{1}}{KT}}+e^{\frac{-E_{2}}{KT}}}
\end{eqnarray}
\subsection{Temperatura del mercado}
Asumiendo que los precios $\mathfrak{p}$, las tasas de cambio de precio (retornos) $\mathfrak{r}$ y los vol\'{u}menes de negociaci\'{o}n $\mathfrak{v}$ jueguen los papeles de posici\'{o}n, velocidad y masa de part\'{i}cula, respectivamente, el Principio de incertidumbre podr\'{i}a expresarse como:
\begin{equation}
    \triangle \mathfrak{p}\triangle (\mathfrak{v}\mathfrak{r})\geq\hbar/2,
\end{equation}
Siendo formalmente $\mathfrak{v}\mathfrak{r}= momentum$ y $\mathfrak{r}=\text{d}\mathfrak{p}/\text{d}t$.
La energ\'{i}a cinetica $E_{k}$ promediada a lo largo del tiempo ser\'{i}a:
\begin{equation}
    \langle E_{k} \rangle=\langle \frac{(\mathfrak{vr})^{2}}{2\mathfrak{v}} \rangle=\frac{1}{2}KT,
\end{equation}
Donde $K$ es la constante de Boltzmann.
Consecuentemente, una coherente definici\'{o}n de temperatura ser\'{i}a:
\begin{equation}\label{Ts}
    T=\frac{\langle\mathfrak{v}\mathfrak{r}^{2}\rangle}{K},
\end{equation}
Para una discusi\'{o}n m\'{a}s amplia del concepto de temperatura del mercado, v\'{e}ase\cite{Subi13}.
\subsection{Distribuci\'{o}n de probabilidad del observable "precio"}
Ahora busquemos la probabilidad $\rho(\mathfrak{p})\text{d}\mathfrak{p}$ de encontrar la part\'{i}cula en una posici\'{o}n $\mathfrak{p}$ situada entre $\mathfrak{p}$ y $\mathfrak{p}+\text{d}\mathfrak{p}$. Cuando la part\'{i}cula est\'{e} en el estado estacionario $|\phi_{n}\rangle$ la correspondiente densidad de probabilidad $\rho_{n}(\mathfrak{p})$ ser\'{a}:
\begin{equation}\label{formRonS}
    \rho_{n}(\mathfrak{p})=|\phi_{n}(\mathfrak{p})|^{2}=\langle\mathfrak{p}|\phi_{n}\rangle\langle\phi_{n}|\mathfrak{p}\rangle,
\end{equation}
En el equilibrio termodin\'{a}mico, el estado de la part\'{i}cula est\'{a} descrito por una colectividad estad\'{i}stica de estados $|\phi_{n}\rangle$ con coeficientes de ponderaci\'{o}n $(1/Z)\exp(-En/(KT))$. Por tanto, la densidad de probabilidad en este caso ser\'{a}:
\begin{equation}\label{formRoS}
    \rho(\mathfrak{p})=\frac{1}{Z}\sum_{n}\rho_{n}(\mathfrak{p})e^{\frac{-E_{n}}{KT}},
\end{equation}
 es decir, podemos definir la densidad de probabilidad $\rho(\mathfrak{p})$ como la suma ponderada de las densidades $\rho_{n}(\mathfrak{p})$ correspondientes a los diversos estados $|\phi_{n}\rangle$.
Veamos c\'{o}mo se relaciona la densidad de probabilidad definida en ~(\ref{formRoS}) con la matriz de densidad $\rho$. Combinando ~(\ref{formRoS}) y ~(\ref{formRonS}) tenemos:
\begin{equation}\label{ropS}
    \rho(\mathfrak{p})=\frac{1}{Z}\sum_{n}\rho_{n}(\mathfrak{p})e^{\frac{-E_{n}}{KT}}\langle\mathfrak{p}|\phi_{n}\rangle\langle\phi_{n}|\mathfrak{p}\rangle,
\end{equation}
Teniendo en cuenta la relaci\'{o}n de clausura para los estados $|\phi_{n}\rangle$, el operador $\exp(-H/(KT))$ puede ser reescrito del modo siguiente:
\begin{equation}\label{eHS}
    e^{\frac{-H}{KT}}=e^{\frac{-H}{KT}}\sum_{n}|\phi_{n}\rangle\langle\phi_{n}|=\sum_{n}e^{\frac{-E_{n}}{KT}}|\phi_{n}\rangle\langle\phi_{n}|,
\end{equation}
Combinando ~(\ref{ropS}) y ~(\ref{eHS}) tenemos que:
\begin{equation}\label{Gs}
    \rho(\mathfrak{p})=\frac{1}{Z}\langle \mathfrak{p}|e^{\frac{-H}{KT}}|\mathfrak{p}\rangle=\langle \mathfrak{p}|\rho|\mathfrak{p}\rangle,
\end{equation}
De este modo, se puede interpretar que $\rho(\mathfrak{p})$ es el elemento diagonal de $\rho$ correspondiente al ket $|{\kern 1pt}\mathfrak{p}\rangle$.
\section{Un ejemplo pr\'{a}ctico descrito paso a paso }
A continuaci\'{o}n, calcularemos la distribuci\'{o}n de probabilidad de precios de un blue-chip perteneciente al selectivo espa\~{n}ol IBEX 35. Los pasos a seguir se describen a continuaci\'{o}n.
\subsection{Fijando las condiciones de contorno}
Lo primero de todo, fijamos el periodo de tiempo durante el cual queremos que sea v\'{a}lido el resultado (distribuci\'{o}n de probabilidad) que obtengamos. Las paredes del pozo determinan un potencial infinito, es decir, la part\'{i}cula cu\'{a}ntica no puede escapar del pozo en el que se encuentra confinada. Esta asunci\'{o}n se basa en la evidencia emp\'{i}rica de que la cotizaci\'{o}n de cualquier valor burs\'{a}til, sea blue-chip o no, no puede superar estad\'{i}sticamente una tasa de cambio m\'{a}xima que depende del periodo de tiempo considerado. As\'{i}, por ejemplo, es muy improbable, salvo crash burs\'{a}til, que un stock oscile por encima del $30$\% dentro de un per\'{i}odo de veinte sesiones de negociaci\'{o}n. Esta limitaci\'{o}n es la que equivale a las paredes del pozo. Consecuentemente, analizando las cotizaciones hist\'{o}ricas del blue chip, determinamos la m\'{a}xima fluctuaci\'{o}n de precios que podr\'{i}a producirse en el periodo de tiempo fijado, con un $95$\% de confianza, salvo procesos de innovaci\'{o}n y crash burs\'{a}til. En el ejemplo que nos ocupa, se fijaron 30 dias como periodo de tiempo, lo que determin\'{o} $\pm15$\% de fluctuaci\'{o}n m\'{a}xima de precios. Lo que implica que las paredes del pozo estar\'{a}n situadas a $\pm15$\% de la \'{u}ltima cotizaci\'{o}n que ocupar\'{a} la posici\'{o}n central.


\begin{figure}[tb]
\centering
 \includegraphics*[scale=.7]{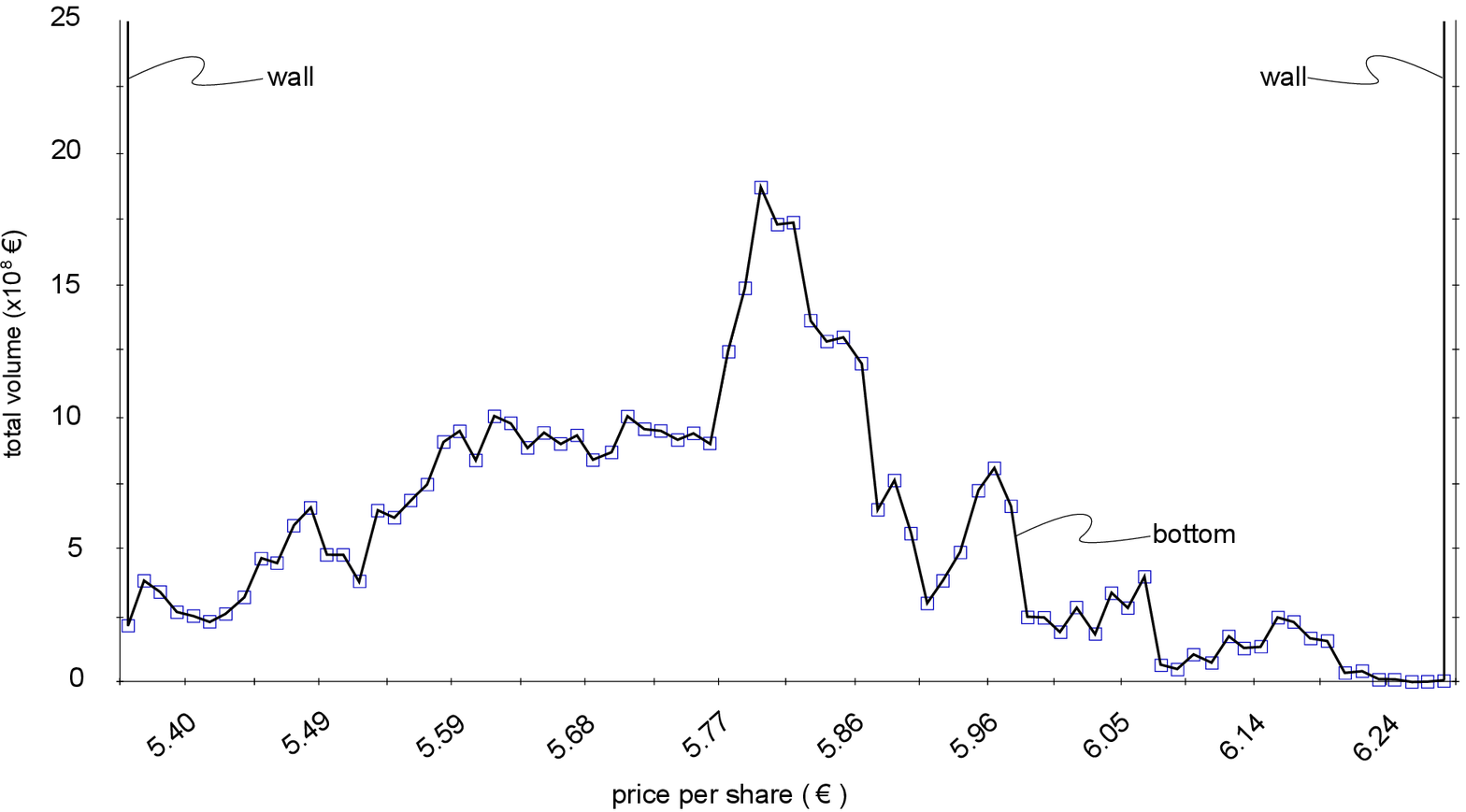}
  \caption{\label{pozoPotS}El campo de potencial, a priori calculado, conforma el fondo del pozo de potencial en el que est\'{a} confinada la part\'{i}cula cu\'{a}ntica}
 \end{figure}
\subsection{Conformando el fondo del pozo}
Aplicando m\'{e}todos de la F\'{i}sica Estad\'{i}stica estimamos la funci\'{o}n $V(\mathfrak{p})$ que tendr\'{a} en general "monta\~{n}as" y "valles" que depender\'{a}n de la historia reciente del mercado. Como ya se\~{n}alamos anteriormente, la descripci\'{o}n de este m\'{e}todo ser\'{a} objeto de un pr\'{o}ximo art\'{i}culo. En la Fig.~\ref{pozoPotS} puede verse el pozo de potencial y la funci\'{o}n $V(\mathfrak{p})$ tal como era estimada a fecha 2 de octubre del 2017. Esta funci\'{o}n $V(\mathfrak{p})$ cambia paulatinamente a lo largo del tiempo, pero para un periodo de unas veinte sesiones de negociaci\'{o}n burs\'{a}til puede considerarse constante, salvo abruptos procesos de innovaci\'{o}n.
\subsection{Resolviendo la ecuaci\'{o}n de Schr\"{o}dinger}
Discretizamos la funci\'{o}n $V(\mathfrak{p})$ en unos $N=100$ puntos separados por un intervalo constante $\triangle=\mathfrak{p}_{i+1}-\mathfrak{p}_{i}$. Con esta discretizacion y aplicando el m\'{e}todo de diferencias finitas, la ecuaci\'{o}n de Schr\"{o}dinger queda reducida a un sistema lineal homog\'{e}neo de $N$ ecuaciones algebraicas que en definitiva es un problema de autovalores de la matriz
\[
\left(
  \begin{array}{ccccc}
    v_{1} & -1 & 0 & \ldots & 0 \\
    -1 & v_{2} & -1 &  & 0 \\
     0& -1 & \ddots & \ddots & \vdots \\
     \vdots&  & \ddots & v_{n-1} & -1 \\
     0& 0 & \ldots & -1 & v_{n} \\
  \end{array}
\right)
\]
 donde $v_{i}=2+\triangle^{2}2\mathfrak{v}V(\mathfrak{p}_{i})/\hbar^{2}$,
 la cual se puede diagonalizar f\'{a}cil y r\'{a}pidamente, quedando el problema resuelto. Se puede usar una implementaci\'{o}n standard sin m\'{a}s que establecer los apropiados factores de conversi\'{o}n de ratio de retorno de precios a distancia en amstrong y de volumen total de negociaci\'{o}n a energ\'{i}a en electron-voltio.
\subsection{Calculando los elementos de la matriz densidad}
El apartado anterior nos habr\'{a} proporcionado los niveles de energ\'{i}a $E_{1}, E_{2},\ldots, E_{n}$, as\'{i} como las funciones de onda estacionarias $\phi_{1}, \phi_{2},\ldots, \phi_{n}$ (ver Fig.~\ref{FO1s}). La temperatura calculada a partir de la volatilidad por la f\'{o}rmula~(\ref{Ts}), junto con los niveles de energ\'{i}a $E_{1}$, $E_{2}$, nos proporcionar\'{a} $\rho_{1,1}$, $\rho_{2,2}$ a partir de las f\'{o}rmulas~(\ref{RonnS}), ~(\ref{RonmS}).
\subsection{Calculando la distribuci\'{o}n de probabilidad de precios}
\begin{figure}[tb]
\centering
 \includegraphics*[scale=.7]{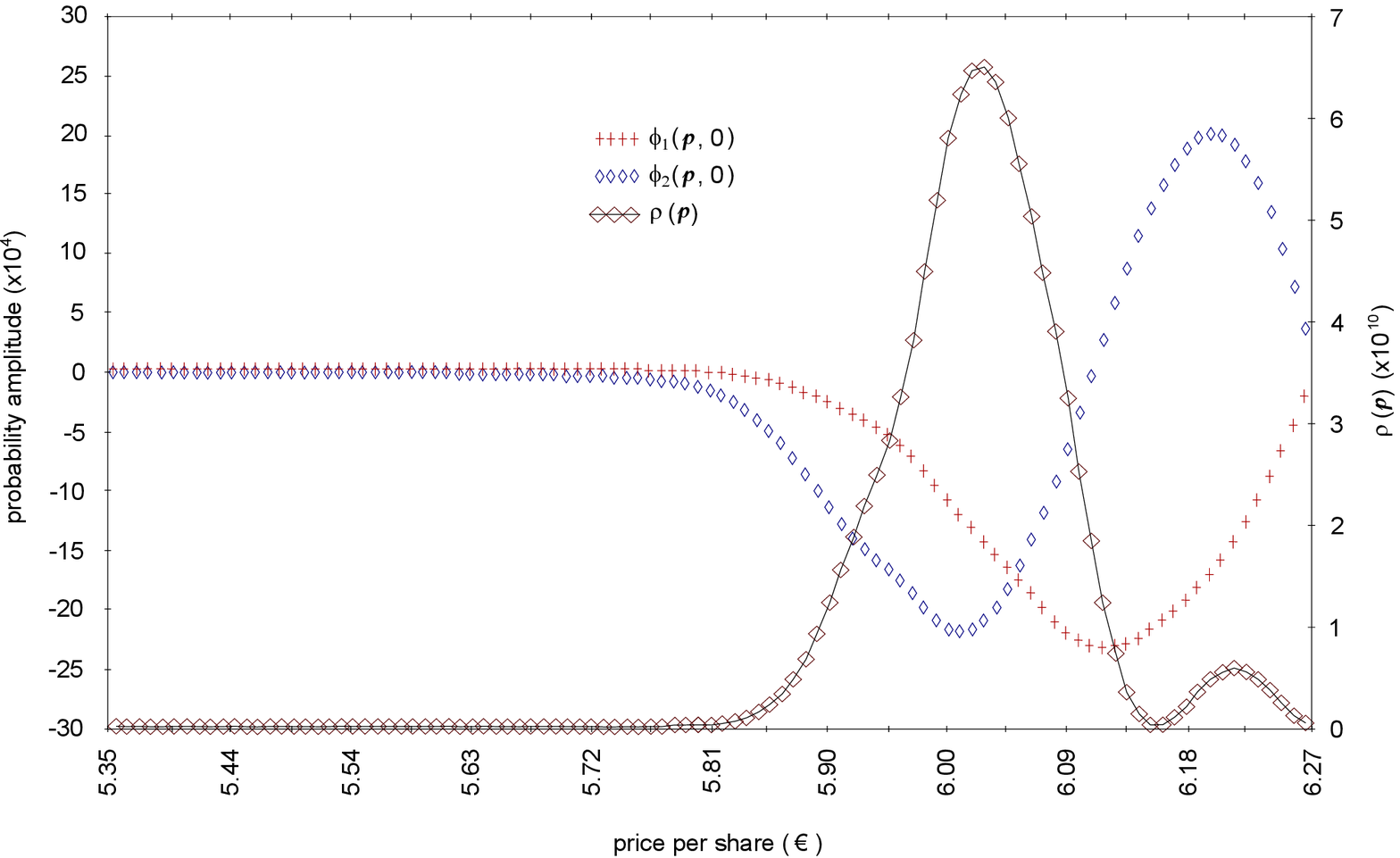}
  \caption{\label{FO1s}Funciones de onda de los dos niveles de energ\'{i}a y distribucion de probabilidad del observable "precio"}
 \end{figure}

La soluci\'{o}n final ser\'{a} la funci\'{o}n $\rho(\mathfrak{p})$ calculada a partir de ~(\ref{Gs}). V\'{e}ase Fig.~\ref{FO1s}.


\section{Conclusiones}
En este art\'{i}culo hemos presentado un modelo para la descripci\'{o}n de la din\'{a}mica de mercados financieros, basado en la evoluci\'{o}n temporal de una part\'{i}cula sometida a un campo de potencial. M\'{a}s concretamente, hemos usado una part\'{i}cula cu\'{a}ntica encerrada en un pozo de potencial infinito, cuyas caracter\'{i}sticas clave son las siguientes:
\begin{itemize}
  \item Respecto a la fluctuaci\'{o}n de precios, se considera el l\'{i}mite estad\'{i}stico que existe de facto en todos los mercados, en lugar del l\'{i}mite legal normativo considerado por otros modelos.
  \item Se considera un pozo infinito de potencial, cuyo fondo no es plano sino funci\'{o}n de la historia reciente del mercado, en lugar del pozo cuadrado infinito, campo externo peri\'{o}dico o potencial harm\'{o}nico de otros modelos.
\end{itemize}
Por nuestra experiencia y hasta el momento, el modelo se muestra coherente y capaz de explicar aspectos de los mercados financieros hasta ahora inexplicados por los modelos basados en part\'{i}culas cl\'{a}sicas (brownianas). Mediante este modelo pueden obtenerse distribuciones de probabilidad de precios para los blue-chips del mercado y usarse con fines predictivos.

\bibliography{SpinModel}

\end{document}